\documentclass[acus]{JAC2000}

\usepackage{graphicx}

\newcommand{\ee}{{\rm e^+e^-}}
\newcommand{\mm}{ \mu^+\mu^-}
\newcommand{\lum}{\cal{L} {\rm }}

\newcommand{\Lunits}{\,{\rm cm^{-2}.s^{-1}}}
\newcommand{\ECoM}{{\rm E_{CoM}}}

\begin{document}
\title{
FURTHER STUDIES ON THE PROSPECTS FOR MANY-TEV MUON COLLIDERS
\thanks{Submitted to Proc. EPAC 2000.
This work was performed under the auspices of
the U.S. Department of Energy under contract no. DE-AC02-98CH10886.}
}

\author{B.J. King, BNL, Upton, NY 11973, USA}

\maketitle

\begin{abstract}
  New self-consistent parameter sets are presented and discussed for
muon collider rings at center-of-mass energies of 10, 30 and 100 TeV.
All three parameter sets attain luminosities of
$\lum = {\rm 3 \times 10^{35}}\;\Lunits$. The parameter sets benefit
from new insights gained at the
HEMC'99 workshop~\cite{hemc99} that considered the feasibility
of many-TeV muon colliders.
\end{abstract}


\section{INTRODUCTION}

   Table 1 of this paper presents self-consistent parameter sets for
muon collider storage rings at center-of-mass energies of $\ECoM = 10$,
30 and 100 TeV. The parameter sets have benefitted and evolved from previous
attempts at defining plausible parameter sets for many-TeV muon colliders.
It is helpful to begin by reviewing these previous studies and their
motivation in order to provide a context for the discussion of the
current parameters.

Parameter sets for muon collider rings at energies up to
$\ECoM=100$ TeV were presented in 1998~\cite{epac98} and 1999~\cite{pac99}.
Following this, a much improved
level of understanding was then obtained from the first
substantial dedicated study of such many-TeV muon colliders, which took
place at the week-long HEMC'99 workshop~\cite{hemc99}.
The majority of the studies at HEMC'99 either assumed or critiqued
straw-man parameter sets~\cite{hemc99specs}, one at $\ECoM=10$ and two
at 100 TeV, that were provided expressly for this purpose.

  Besides presenting an overview of the HEMC'99 parameter sets,
reference~\cite{hemc99specs}
also reviewed the feed-back on the parameters that was provided by the
workshop. This paper should be referred to for many discussions that
remain relevant for the current parameter sets of table 1.

  The 48 participants
at HEMC'99 considered side-by-side the accelerator challenges and
the high energy physics (HEP) potential of many-TeV muon colliders.
The HEP motivation for the workshop was very strong because
experimental discoveries in HEP normally come
from advances in energy reach, as has been emphasized and discussed in,
for example,
references~\cite{Willis} and~\cite{hemc99intro}. HEP discussions specific
to many-TeV muon colliders can be found in~\cite{pr98} and, mainly,
in the HEMC'99 Proceedings~\cite{hemc99}.

  Of the three many-TeV parameter sets in table 1, those at
10 TeV and 100 TeV evolved
directly from the corresponding 10 TeV and (the first of the) 100 TeV
parameter sets for HEMC'99, taking into account the constructive
criticisms that emerged from the workshop.
A mid-point energy was
considered valuable for examining parameter trends with increasing
energy, and the 30 TeV parameter set provides such an interpolation
between the lower and higher energy sets.

 Invaluable benchmarks for all of these many-TeV studies were
provided by lower energy parameters
that have been studied and evaluated~\cite{Snowmass,status} by the
Muon Collider Collaboration (MCC).
The first column of table 1 shows, for comparison, the range of
parameters for the muon colliders in the range $\ECoM=0.1$, 3 TeV
from the MCC's status report~\cite{status}.

\section{DISCUSSION ON PARAMETER SETS} 

  The energy scale and some other parameter choices in table 1
were strongly influenced by considerations of synchrotron radiation.
This imposes a natural cut-off scale for circular
muon storage rings in the range $\ECoM \sim 100$ TeV since
the synchrotron radiation loss at such energies has risen
rapidly to become comparable to
the beam power. At HEMC'99, Telnov made the additional
observation~\cite{Telnov}
that the
quantum nature of the sychrotron radiation could lead to beam
heating, rather than cooling, for sufficiently high beam energies and
small emittances. This observation effectively invalidated the more
aggressive of the two HEMC'99 parameter sets at 100 TeV -- which
therefore won't
be discussed further in this paper -- and also cast
some doubt on the 100 TeV parameter set with the larger emittance.

  The synchrotron radiation concerns were addressed in the 100 TeV
parameter set in table 1 by:
\begin{enumerate}
  \item  raising the emittance in each of the transverse coordinates
by the large factor of 90. This should comfortably address Telnov's
concern and result in net synchrotron cooling by raising the horizontal
emittance to well above the quantum break-even value
\item  increasing
the collider ring circumference by a factor of two and, correspondingly,
reducing the average bending magnetic field by a factor of two, to
5.3 Tesla
\item reducing the average beam current by nearly a factor of 2, to 4 mA.
\end{enumerate}
The combined effect of the second and third changes was to reduce the
synchrotron radiation to 50 MW, down from the previous,
somewhat problematic level of
195 MW in the HEMC'99 parameter set. Although still a factor of 2.5 larger
than the synchrotron power at LEP II, this reduced level was considered very
appropriate for a far future collider at the energy frontier.

 These changes
should also help to address reservations expressed by Harrison~\cite{Harrison}
at HEMC'99 about the feasibility of 10 Tesla cosine theta dipoles in the
presence of large amounts of synchrotron radiation. Besides lowering the
average required magnetic field by a factor of two, it is noted that
the synchrotron radiation power deposited per unit length around the
collider ring has fallen by almost a factor of 8 from the HEMC'99
parameter set at 100 TeV.

  In addition to the adjustments just mentioned that were specific
to the 100 TeV parameter set, all three many-TeV parameter sets
in table 1 were made more conservative than the HEMC'99 parameter
sets in several areas:
\begin{itemize}
  \item  in recognition of the difficulty and novelty of ionization cooling,
the phase space densities in table 1 were all scaled back to coincide
with the upper end of the parameter choices from reference~\cite{status}
for lower energy muon colliders, i.e. $2.4 \times 10^{22}$ ${\rm m}^{-3}$.
  \item  the final focus parameters are perhaps the most difficult
of all for a non-specialist to evaluate. As has been discussed in
references~\cite{epac98,pac99,hemc99specs}, the final focus difficulty
can be usefully benchmarked to other muon collider and $\ee$
collider parameter
sets according to the value of 3 parameters in particular: the
$\beta^*$ in the x and y coordinates
and of two other defined
parameters, the so-called ``demagnification factor'' and
``chromaticity quality factor''. All three benchmark parameters
have been somewhat relaxed in response to feed-back~\cite{hemc99specs}
from the studies by final focus lattice experts at HEMC'99. Further
explicit magnet lattice designs, now for each of the three parameter
sets in table 1, would be invaluable for assessing whether the new,
more relaxed parameters have reached an acceptable level of plausibility
  \item  the average beam currents and resulting beam powers were reduced
so that the worst case, at 100 TeV, had a summed beam plus synchrotron
power of 180 MW, i.e. comparable to the 170 MW beam power that has been
under consideration for the Accelerator Production of Tritium
project~\cite{APT}
  \item  the beam-beam tune disruption parameter was
lowered slightly for all three sets to a value, in the worst case,
of $\Delta \nu = 0.091$. This is not far above the
impressive new LEP II record of $\Delta \nu = 0.083$
that was reported in this
conference~\cite{LEPIItuneshift}.
\end{itemize}

  The unavoidable cost of these relaxed machine parameters was to lower 
the luminosity to $\lum = {\rm 3 \times 10^{35}}$ $\Lunits$ for each of
the 10, 30 and 100 TeV parameters. This is a reduction to
30\% of the luminosities,
$\lum = {\rm 1 \times 10^{36}}$ $\Lunits$,
of the corresponding HEMC'99 parameter sets for 10 TeV and 100 TeV.
To put this in perspective, the new luminosities are still orders
of magnitude higher than at any existing colliders and are also higher 
than any speculated parameters the author is aware of for plausible
future machines other than muon colliders.

\section{SUMMARY}

 The extremely high constituent particle energies and luminosities
of the parameter sets presented in table 1
continue to emphasize the impressive potential of muon colliders
for exploring the energy frontier of elementary particle physics.
Therefore, further paper studies and simulations for many-TeV muon
colliders should continue to play a valuable role in our field.
More specifically, the parameter sets presented in this paper would
certainly benefit from feed-back and constructive criticism by experts
in areas such as the design of final focus lattices.

\begin{table*}[htb]
\centering
\renewcommand\tabcolsep{5pt}
\caption{
Self-consistent collider ring parameter sets for many-TeV muon colliders.
For comparison, the first column displays the range of parameters for the
lower energy muon colliders discussed in reference~\cite{status}.}
\begin{tabular}{|r|cccc|}
\hline
\multicolumn{1}{|c|}{ {\bf parameter set} }
                            &     & A  & B  & C \\
\multicolumn{1}{|c|}{ {\bf center of mass energy, ${\rm E_{CoM}}$} }
                            & 0.1 to 3 TeV & 10 TeV  & 30 TeV  &  100 TeV \\
\hline \hline
\multicolumn{1}{|l|}{\bf collider physics parameters:} & & & & \\
luminosity, ${\cal L}$ [${\rm 10^{35}\: cm^{-2}.s^{-1}}$]
                                        & $8 \times 10^{-5}$$\rightarrow$0.5
                                        & 3.0 & 3.0 & 3.0 \\
$\int {\cal L}$dt [${\rm fb^{-1}/year}$]
                                        & 0.08$\rightarrow$540
                                        & 3000 & 3000 & 3000 \\
No. of $\mu\mu \rightarrow {\rm ee}$ events/det/year
                                        & 650$\rightarrow$10 000
                                        & 2600 & 290 & 26 \\
No. of 100 GeV SM Higgs/year            & 4000$\rightarrow$600 000
                                        & $4 \times 10^6$
                                        & $5 \times 10^6$
                                        & $6 \times 10^6$ \\
CoM energy spread, ${\rm \sigma_E/E}$ [$10^{-3}$]
                                        & 0.02$\rightarrow$1.1 & 0.42 & 0.080 & 0.071 \\
\hline
\multicolumn{1}{|l|}{\bf collider ring parameters:}   & & & & \\
circumference, C [km]                   & 0.35$\rightarrow$6.0
                                        & 15 & 39 & 200 \\
ave. bending B field [T]                & 3.0$\rightarrow$5.2
                                        & 7.0 & 8.1 & 5.2 \\
\hline
\multicolumn{1}{|l|}{\bf beam parameters:}            & & & & \\
($\mu^-$ or) $\mu^+$/bunch, ${\rm N_0[10^{12}}]$
                                        & 2.0$\rightarrow$4.0
                                        & 2.9 & 2.0 & 1.6 \\
($\mu^-$ or) $\mu^+$ bunch rep. rate, ${\rm f_b}$ [Hz]
                                        & 15$\rightarrow$30
                                        & 15 & 7.5 & 5 \\
6-dim. norm. emit., $\epsilon_{6N}
               [10^{-12}{\rm m}^3$]     & 170$\rightarrow$170
                                        & 125 & 85 & 70\\
$\epsilon_{6N}
   [10^{-4}{\rm m}^3.{\rm MeV/c}^3$]
                                        & 2.0$\rightarrow$2.0
                                        & 1.5 & 1.0 & 0.83 \\
P.S. density, ${\rm N_0}/\epsilon_{6N}
               [10^{22}{\rm m}^{-3}$]   & 1.2$\rightarrow$2.4
                                        & 2.3 & 2.4 & 2.3 \\
x,y emit. (unnorm.)
              [${\rm \pi.\mu m.mrad}$]  & 3.5$\rightarrow$620
                                        & 0.84 & 0.19 & 0.040 \\
x,y normalized emit.
              [${\rm \pi.mm.mrad}$]     & 50$\rightarrow$290
                                        & 40 & 27 & 19 \\
long. emittance [${\rm 10^{-3}eV.s}$]   & $0.81\rightarrow24$
                                        & 28 & 40 & 68 \\ 
fract. mom. spread, $\delta$ [$10^{-3}$]
                                        & 0.030$\rightarrow$1.6
                                        & 0.50 & 0.20 & 0.075 \\
relativistic $\gamma$ factor, ${\rm E_\mu/m_\mu}$
                                        & 473$\rightarrow$14 200
                                        & 47 300 & 142 000 & 473 000 \\
time to beam dump,
          ${\rm t_D} [\gamma \tau_\mu]$ & no dump
                                        & no dump & no dump & no dump \\
effective turns/bunch                   & 450$\rightarrow$780
                                        & 1040 & 1200 & 780 \\
ave. current [mA]                       & 17$\rightarrow$30
                                        & 29 & 12 & 4.0 \\
beam power [MW]                         & 1.0$\rightarrow$29
                                        & 70 & 72 & 128 \\
synch. rad. critical E [MeV]            & $5 \times 10^{-7} \rightarrow
                                                           8 \times 10^{-4}$
                                        & 0.012 & 0.12 & 1.75 \\
synch. rad. E loss/turn [GeV]           & $7 \times 10^{-9} \rightarrow
                                                           3 \times 10^{-4}$
                                        & 0.017 & 0.52 & 25 \\
synch. rad. power [MW]                  & $1\times10^{-7}\rightarrow$0.010
                                        & 0.48 & 6.0 & 50 \\
beam + synch. power [MW]                & 1.0$\rightarrow$29
                                        & 70 & 78 & 180 \\
power density into magnet liner [kW/m]  & 1.0$\rightarrow$1.7
                                        & 2.0 & 0.84 & 0.48 \\
\hline
\multicolumn{1}{|l|}{\bf interaction point parameters:}      & & & & \\
spot size, $\sigma_{x,y}$ $[\mu {\rm m}]$
                                        & 3.3$\rightarrow$290
                                        & 1.7 & 0.88 & 0.47 \\
bunch length, $\sigma_z$ [mm]           & 3.0$\rightarrow$140
                                        & 3.4 & 4.0 & 5.4 \\
$\beta^*_{x,y}$ [mm]                    & 3.0$\rightarrow$140
                                        & 3.4 & 4.0 & 5.4 \\
ang. divergence, $\sigma_\theta$
                             [mrad]     & 1.1$\rightarrow$2.1
                                        & 0.50 & 0.22 & 0.086 \\
beam-beam tune disruption, $\Delta \nu$
                                        & 0.015$\rightarrow$0.051
                                        & 0.079 & 0.079 & 0.091 \\
pinch enhancement factor, ${\rm H_B}$   & 1.00$\rightarrow$1.01
                                        & 1.06 & 1.06 & 1.09 \\
beamstrahlung frac. E loss/collision    & negligible
                                        & $2.3 \times 10^{-8}$
                                        & $1.0 \times 10^{-7}$
                                        & $5.5 \times 10^{-7}$
                                           \\
\hline
\multicolumn{1}{|l|}{\bf final focus lattice parameters:} & & & & \\
max. poletip field of quads., ${\rm B_{5\sigma}}$ [T]
                                        & 6$\rightarrow$12
                                        & 12 & 12 & 12 \\
max. full aper. of quad., ${\rm A_{\pm5\sigma}}$[cm]
                                        & 14$\rightarrow$24
                                        & 21 & 25 & 31 \\
quad. gradient,    $2{\rm B_{5\sigma} / A_{\pm5\sigma}}$[T/m]
                                        & 50$\rightarrow$90
                                        & 120 & 97 & 77 \\
${\rm \beta_{max} [km]}$                & 1.5$\rightarrow$150
                                        & 520 & 3200 & 24 000 \\
ff demag., $M \equiv \sqrt{\beta_{\rm max}/\beta^*}$
                                        & 220$\rightarrow$7100
                                        & 12 000 & 28 000 & 67 000 \\
chrom. quality factor, $Q \equiv M \cdot \delta$
                                        & 0.007$\rightarrow$11
                                        & 6.2 & 5.7 & 5.0 \\
\hline
\multicolumn{1}{|l|}{\bf neutrino radiation parameters:} & & & & \\
collider reference depth, D[m]     & 10$\rightarrow$300 & 100 & 100 & 100 \\
ave. rad. dose in plane [mSv/yr]   & $2 \times 10^{-5}$$\rightarrow$0.02
                                        & 1.2 & 4.8 & 20 \\
str. sec. len. for 10x ave. rad. [m]
                                   & 1.3$\rightarrow$2.2 & 0.95 & 1.6 & 8.4 \\
$\nu$ beam distance to surface [km]     & 11$\rightarrow$62 & 36 & 36 & 36 \\
$\nu$ beam radius at surface [m]        & 4.4$\rightarrow$24
                                        & 0.75 & 0.25 & 0.075 \\ \hline
\end{tabular}
\label{colliderpara}
\end{table*}

\end{document}